\documentclass[aps,prd,nofootinbib,preprint]{revtex4}
\usepackage{amsmath,amssymb,graphicx,mathrsfs,slashed}
\usepackage[usenames, dvipsnames]{color}
\usepackage{tabularx}
\usepackage[
verbose,
colorlinks=true,
naturalnames=true,
linkcolor=blue,
citecolor=black,
]{hyperref}
\newcommand{\be}{\begin{equation}}
\newcommand{\ee}{\end{equation}}
\newcommand{\bea}{\begin{eqnarray}}
\newcommand{\eea}{\end{eqnarray}}

\def\({\left(} \def\){\right)}

\begin{document}

\title{\Large Quantum Love}

\author{\large Ram Brustein, Yotam Sherf }
\affiliation{\ \\ Department of Physics, Ben-Gurion University,
					Beer-Sheva 84105, Israel \\
ramyb@bgu.ac.il, sherfyo@post.bgu.ac.il}

\begin{abstract}
The response of a gravitating object to an external tidal field is encoded in its Love numbers, which identically vanish for classical black holes (BHs). Here we show, using standard time-independent quantum perturbation theory, that for a quantum BH, generically, the Love numbers are nonvanishing and negative, and that their magnitude depends on the lowest-lying levels of the quantum spectrum of the BH. We calculate the quadrupolar electric quantum Love number of slowly rotating BHs and show that it depends most strongly on the first excited level of the quantum BH. We then compare our results to the same Love number of exotic ultra compact objects and to that of classical compact stars and highlight their different parametric dependence. Finally, we discuss the detectability of the quadrupolar quantum Love number in future precision gravitational-wave observations and show that, under favorable circumstances, its magnitude is large enough to imprint an observable signature on the gravitational waves emitted during the inspiral phase of two moderately spinning BHs.
\end{abstract}
\maketitle
%%\renewcommand{\baselinestretch}{1}\normalsize
%%
%%\newpage
%%
%\maketitle	
		\section{Introduction}
\begin{subequations}
	\renewcommand{\theequation}{\theparentequation.\arabic{equation}}

The gravitational-wave (GW) observations by LIGO and the future observations by the planned Laser Interferometer Space Antenna (LISA), offer opportunities for testing strong gravity effects through precision GW measurements during the inspiral phase of a compact binary system \cite{LISA,TheLIGOScientific:2017qsa,Abbott:2020uma,AmaroSeoane:2012je}.  As the two companions spiral around each other, they are tidally deformed \cite{Thorne:1980ru,Finn:2000sy}, leaving a specific imprint on the emitted GW waveform \cite{Flanagan:1997sx,Flanagan:2007ix,Hinderer:2009ca,Yagi:2016bkt,Abbott:2018exr,De:2018uhw}. The tidal response of each of the companions is quantified in terms of  the tidal Love numbers.

The weak external tidal field induces, generically, small nonvanishing mass (electric) and current (magnetic) moments. In the linear response approximation, the moments are proportional to the external tidal field.  The largest of these induced moments is typically the mass quadrupole, which is proportional to the quadrupolar tidal field $\mathcal{E}_{ab}$, $Q_{ab}=-\frac{2}{3}k_2R^5\mathcal{E}_{ab}$. Here $k_2$ is the dimensionless quadrupolar electric tidal Love number and $R$ is the radius of the inspiraling object.

The calculation of  $k_2$ is performed in great detail in \cite{Hinderer:2007mb,Damour:2009vw,Binnington:2009bb}. Its value is most sensitive to the object compactness\footnote{We use relativistic units $G$, $c=1$ and consider nonrotating BHs unless stated otherwise.} $C=M/R$. For the case that $C$ approaches that of a  black hole (BH), $C\rightarrow 1/2$, the Love number exhibits a universal decrease, tending precisely to zero in the BH limit. This universal behavior is a consequence of the BH no-hair property \cite{Damour:2009vw,Gurlebeck:2015xpa,Porto:2016zng,Yagi:2016bkt}.  The exact vanishing  of $k_2$ for BHs \footnote{In \cite{LeTiec:2020spy,LeTiec:2020bos} it is claimed that the Love number for spinning BHs in an axisymmetric tidal field $(m\neq 0)$ is nonvanishing. The results were challenged in \cite{Chia:2020yla}.} and being the largest of the dimensionless Love numbers,  makes $k_2$ a key diagnostic for any deviations from classical general relativity (GR).

In \cite{Cardoso:2017cfl}, the Love numbers for several exotic ultracompact objects (UCOs) were calculated and were shown not to vanish. The numerical results exhibit a universal, model-independent, logarithmic suppression on the relative deviations from the Schwarzschild radius  $R=2M(1+\epsilon)$.

We are interested in calculating the Love numbers of large astrophysical BHs. As for any macroscopic object,  the Bohr correspondence principle implies that some quantum state corresponds to the classical BH, no matter how large it is. In the following, we use the term ``quantum black hole" (QBH) to mean the quantum state that corresponds to a classical BH. The QBH is therefore a UCO that possesses a horizon and, in addition, has a discrete spectrum of quantum mechanical energy levels. These energy levels can be viewed as coherent states that correspond to macroscopic, semiclassical excitations of the QBH. In the ground state of the QBH, the exterior geometry is exactly the Schwarzschild geometry. But, when a QBH is in an excited state, it displays deviations from its GR description, and therefore it can be, in principle, distinguished from its classical counterpart.

The classical BH is bald, while the QBH has some  ``quantum hair" \cite{Brustein:2017nis,Kourkoulou:2017zaj,Brustein:2018fkr}. Moreover, the properties of the quantum hair can be entirely explained by an external observer via the Bohr correspondence principle that requires some specific changes to the near horizon geometry, without any need to invoke  new physical principles \cite{Brustein:2017nis,Brustein:2017koc}.  The amount of information that the quantum hair carries is limited. However, if observed, it could provide  unrivaled information on some properties of the spectrum of the QBH \cite{Brustein:2017koc,Giddings:2014nla,Giddings:2017mym,Brustein:2019twi,Sherf:2019arn}. Quantum imprints due to tidal heating in the inspiral phase were also studied recently in \cite{Datta:2020rvo,Agullo:2020hxe}.

We will show that the Love numbers are part of this quantum hair and can, in principle, be observed. In practice, it is $k_2$ that seems to  offer the best opportunity for detection.

Quantum effects for large astrophysical BHs are universally expected to be negligibly small, based on the expectation that the strength of quantum effects is controlled by the extremely small ratio of the Planck length squared to typical curvatures $l_P^2/R_S^2$. However, we argue in the following  that the strength of quantum effects for QBHs can be much larger.

In GR, the interior of a BH is empty except for a possibly singular core.  The firewall argument marked the beginning of a new era in the theory of QBHs \cite{AMPS,MP}, indicating that this picture is in need of a substantial revision. Forerunners of the argument and a more recent review can be found in \cite{Sunny,Mathur1,Braun,Mathur2017}, respectively.

Putting remnants aside, two main classes of solutions to the firewall problem emerged as possible candidates.  In the first class the horizon region is a vacuum, but novel nonlocal physics is introduced to resolve the information paradox: the degrees of freedom very far from the horizon are not distinct from the degrees of freedom inside the horizon \cite{erepr,newmalda}.   The singularity is often viewed as irrelevant, under the premise that it will be regularized somehow in a way that does not affect the structure of spacetime on horizon scales.

In the second class,  BHs are described by nonsingular states that do not collapse under their own gravity. Strong quantum effects  ``smear" the would-be singularity over horizon-sized length scales.  These changes lead to a spectrum of excitations whose characteristic scale is the horizon rather than the Planck length. The self-consistency of this description of the interior requires a significant departure from semiclassical gravity, as well as some exotic matter which is outside the realm of the standard model \cite{BHFollies}. Fuzzballs \cite{MathurFB,otherfuzzball} and the polymer model \cite{strungout} are in this class. The new physics that resolves the singularity introduces a new scale, and in addition to the Planck scale, the ratio of the two scales can be viewed as a coupling constant.  For example, in string theory, this length scale is the string scale $l_s$, and it is rather the ratio $g_s^2=l_P^2/l_s^2$ that controls the strength of quantum effects. The magnitude of $g_s^2$ is expected to be small, but of the order of all other known gauge couplings, $ g_s^2 \sim 0.1$.

Here, we present a general, closed expression for both electric (polar) and magnetic (axial) Love numbers (tensor) for  QBHs in terms of their spectrum.    The calculation is performed in an analogy to the calculation of the polarizability of an atom by using second-order time-independent perturbation theory. We show that the Love numbers are most sensitive to the lowest-lying energy level. From this perspective, the Love numbers do not vanish because the tidal field mixes a small amount of the first excited level with the ground state.

In a follow up paper \cite{Brustein:2021bnw}, we describe explicitly the connection between the classical and quantum Love calculations using the ideas presented in \cite{Lai:1993di,Ho:1998hq,Chakrabarti:2013xza,Andersson:2019ahb}. We first establish an effective description for the interior fluid modes of ultracompact objects as a collection of driven harmonic oscillators characterized  by their frequencies. We then find the appropriate boundary conditions on the perturbed Einstein equations and show that derivation of the quantum Love number of a quantum black hole matches exactly the standard classical calculation of the Love number \cite{Hinderer:2007mb,Damour:2009vw,Binnington:2009bb}, when quantum expectation values are replaced by the corresponding classical quantities, as dictated by the Bohr correspondence principle. The quantum Love number is equal to the classical Love number that is computed in the traditional way.
The current paper and \cite{Brustein:2021bnw} have different goals. The goal of the current paper  is to study the response of a general quantum system to an external tidal field and demonstrate how it acquires nonvanishing Love numbers.  On the other hand, the motivation of \cite{Brustein:2021bnw} is to demonstrate how an object that possesses a horizon can have a nonvanishing Love number.  They are similar in that both rely on the interpretation of the nonrelativistic fluid modes as large quantum excitations.

The paper is organized as follows.  In Sec.\ref{int} we review the standard  calculation of the atom's electric polarizability using time-independent perturbation theory. Then, by replacing the external electric field and the dipole moment with the gravitational tidal field and the mass and current moments, respectively, we derive a general expression for the gravitational polarizability of a quantum mechanical object---the Love numbers. Next, in Sec.\ref{calc}, by applying the Bohr correspondence principle we evaluate the Love number and find that it is negative, and its magnitude depends on the lowest-lying levels of the quantum spectrum of the QBH.
We demonstrate the ideas by replacing the large excitations spectrum of the QBH with an analogous semiclassical fluidlike description. Then by imposing generic boundary conditions, we provide an explicit expression of the Love number of QBHs.
Finally, in Sec.\ref{dec} we discuss the possible observation of the quantum Love numbers. We show that, under favorable circumstances,  future LISA observations could indeed detect them by precise measurement of the spectrum of GWs emitted during the inspiral phase of a binary system of supermassive moderately spinning BHs.
In the Appendix, we discuss the promotion of the magnetic Love numbers of a slowly rotating object to tensors and the spin corrections to the tidal Love numbers.

	\end{subequations}
\section{{ Quantum Love numbers}}\label{int}
	\begin{subequations}
	\renewcommand{\theequation}{\theparentequation.\arabic{equation}}
				
As a prelude to the	calculation of the quantum Love numbers, we briefly recall the analogous calculation of the polarizability of an atom.  The atom is placed in a region of an approximately uniform electric field $\mathcal{E}_i$ that is induced by a weak external potential $U_{ext}$, $\mathcal{E}_i=-\frac{\partial U_{ext}}{\partial x^i}$.  The interaction of the atom with the external electric field, is expressed in terms of the dipole moment $D=\int\rho(x^{'})x^{'}dV^{'}$, where the integral is performed over the charge distribution. The interaction is given by  $\widehat{V}_{int}~=~-\mathcal{E}_i\widehat{D}_i$.
The induced dipole moment of the perturbed atom can be calculated in second-order time-independent perturbation theory \cite{sakurai}. In this case, symmetry implies that the atom's linear response to the external electric field is then $\langle\Psi_0|\widehat{D}_i|\Psi\rangle=\alpha \mathcal{E}_i$, where $|\Psi_0\rangle=|1,0,0\rangle$ is the ground state of the atom, $|\Psi\rangle$ is the first-order correction to the atom ground state
$
|\Psi\rangle=|\Psi_0\rangle+\sum_{n,l,m}|n,l,m\rangle \frac{\langle 1,0,0|\hat{V}_{int}|n,l,m\rangle }{\Delta E_{1,n}}$, and $\alpha$ is the electric polarizability,
\be
\alpha~=~\sum_{n,|m|\le1}\frac{|\langle \Psi_0|\widehat{D}_i|n,l=1,|m|\le 1\rangle|^2 }{\Delta E_{1,n}}~,
\ee
and where $l$ and $m$ are the angular quantum numbers, $n$ is the radial quantum number and  $\Delta E_{1,n}= E_1-E_n$.

 We derive an expression for the gravitational polarizability---the Love numbers---by replacing the external electric field and the dipole moment by the tidal field and the mass and current moments, respectively.
		
We consider the inspiral phase of a binary system, where one of the companions is an object of mass $M_{ext}$ on a circular orbit of radius $b$ and the other is a  nonrotating QBH of mass $M_{BH}$ and radius $R_S$.  In the early stages of the inspiral, the BH responds to the external slowly varying tidal field that is generated by its companion. For $b\gg R_S$ one can expand the Newtonian potential $U_{ext}~=-{M_{\text{ext}}}/{|\vec{b}-\vec{x}|}$ of the external body in the vicinity of the BH in its local inertial frame,
$
		U(t,x)_{ext}=U_{ext}(0)+\frac{1}{2}\frac{\partial^2 U_{ext}}{\partial x^i \partial x^j}\Big|_0x^{i'}x^{j'} +\cdots.
$

The interaction of the QBH with the external field is expressed in terms of the quantum trace-free symmetric  mass and current multipole moments, $\widehat{Q}^{(l)}$ and $\widehat{S}^{(l)}$, these being the quantum counterparts of the classical multipoles \cite{Thorne:1980ru}. We further assume that the expectation value of the mass and current moments of the BH vanishes in the BH ground state, as dictated by the angular symmetry of the multipole operators and in accordance with the classical no-hair theorems, denoting the ground state of the BH by $|\Psi_0\rangle $,
$\langle\Psi_0|\widehat{Q}^{(l)}|\Psi_0\rangle=0$, $\langle\Psi_0|\widehat{S}^{(l)}|\Psi_0\rangle=0$. Since the external potential is slowly varying, time-independent perturbation theory should be a good approximation.

Let us evaluate explicitly the correction to the ground state energy due to the induced quadrupole, $\widehat{Q}_{ij}$. We follow here the conventions of \cite{Hinderer:2007mb}, in analogy to the electric polarizability calculation,
$
\widehat{V}_{int}=- \frac{1}{2}\mathcal{E}_{ij}\widehat{Q}_{ij}~,
$
where $\mathcal{E}_{ij}=\frac{\partial^2 U_{ext}}{\partial x^i \partial x^j}$ is the tidal field.
The sign of the interaction term is important and leads, generically, to negative quantum Love numbers. For neutron stars, the sign of the interaction term is positive and it leads to positive Love numbers \cite{Hinderer:2007mb,Thorne:1997kt,Cardoso:2017cfl}. The physical reason is that for BHs, the mass as a function of the radius $M(R)$ is an {\em increasing} function, while for neutron stars it is a {\em decreasing} function (see Fig.~2 of \cite{Thorne:1997kt}). For UCOs, the sign of the Love number is also, generically, negative.

The leading-order correction to the BH ground state quadrupole is given by
\be
\langle\Psi_0|\widehat{Q}_{kl}|\Psi\rangle= -\mathcal{E}_{ij}\hspace{-0.2in}\sum_{n_r> 1,|m|\le 2}\hspace{-0.2in}\dfrac{\langle \Psi_0|\widehat{Q}_{ij}|n_r,2,m\rangle \langle n_r,2,m|\widehat{Q}_{kl}|\Psi_0\rangle}{\Delta E_{1,n_r}}~,
\label{qi}
\ee
where $\Delta E_{1,n_r}= E_1-E_{n_r}$. Here the radial number of the ground state $\Psi_0$ is denoted by $n_r=1$, so the energy of the ground state is $E_1=M_{BH}$.
Symmetry implies that the BH electric quadrupolar Love number is  given by
$
\frac{1}{2}\langle\Psi_0|\widehat{Q}_{ij}|\Psi\rangle=-\lambda_2 \mathcal{E}_{ij}~.
\label{lqm}
$
Here $\lambda_2$ is the dimensional quadrupolar Love number. The dimensionless Love number $k_2$ is commonly defined as  $k_2=\frac{3}{2}R^{-5}\lambda_2$. From Eq.~(\ref{qi}), it follows  that	
		\begin{gather}
		k_2~=~-\dfrac{3}{2 R^{5}}\sum_{n_r>1,|m|<2}\dfrac{1}{2} \dfrac{|\langle \Psi_0|\widehat{Q}_{ij}|n_r,2,m\rangle|^2 }{|\Delta E_{1,n_r}|}~.
		\label{k22}
		\end{gather}
Equation~(\ref{k22}) is the main result of our paper. It demonstrates that, generically, a quantum mechanical object must have a nonvanishing quadrupolar Love number that depends solely on the quantum state of the object and its energy spectrum. 	The negative sign of $k_2$ reflects the fact that the energy of a BH increases when its radius becomes larger, as previously explained. \footnote{This argument is also supported by the \textit{shape Love number} \cite{Damour:2009va,pw}.} As it is previously explained, in \cite{Brustein:2021bnw}, we showed explicitly that the quantum Love number is equal to the classical Love number that is computed in the traditional way when quantum expectation values are replaced by the corresponding classical quantities, as dictated by the Bohr correspondence principle.

The general expressions for the higher-$l$ electric and magnetic quantum Love tensors can be obtained  by following  the steps that led to Eq.~(\ref{qi}):
 \begin{gather}
\hspace{-0.6 cm} k^{E}_l\!\!\!~=~-\hspace{-0.2in}\sum_{n_r>1,l,|m|\leq l} \hspace{-0.2in}\dfrac{1}{R^{2l+1}}\dfrac{(2l-1)!!}{2(l-2)!}\dfrac{1}{l!}~ \dfrac{|\langle \Psi_0|\widehat{Q}^{( l)}|n_r,~l,~m\rangle|^2 }{\Delta E_{1,n_r}}~, \\
 \hspace{-0.45 cm}k^{B}_{l}\!\!~=~-\hspace{-0.2in}\sum_{n>1,l,|m|\leq l}\hspace{-0.2in}\dfrac{1}{R^{2l+1}}\dfrac{(l+1)(2l-1)!!}{6(l-2)!}\dfrac{1}{l !}~ \dfrac{|\langle \Psi_0|\widehat{S}^{( l)}|n_r,~l,~m\rangle|^2 }{\Delta E_{1,n_r}}~.
 \label{magk}
 \end{gather}
Recently, in \cite{Poisson:2020mdi} it was shown that the magnetic Love numbers of a slowly rotating object should be promoted to tensors.  We discuss this in more detail in addition to the spin corrections to the tidal Love number in the Appendix.

Again, the conclusion is that, generically, QBHs must posses nonvanishing Love numbers.

	\end{subequations}

\section{{ Electric quadrupolar quantum Love number}}\label{calc}	
\begin{subequations}
\renewcommand{\theequation}{\theparentequation.\arabic{equation}}

The starting point of our evaluation of $k_2$ is Eq.~(\ref{k22}).
The external quadrupole tidal field is proportional to the spherical harmonic $Y_{20}$ due to the symmetry of the inspiral trajectory. The induced quadrupole shares this angular dependence. It follows that
\begin{gather}
		k_2~=~-\dfrac{3}{4 R^{5}}\sum_{n_r} \dfrac{|\langle \Psi_0|\widehat{Q}|n_r,2,0\rangle|^2 }{|\Delta E_{1,n_r}|}~.
		\label{k22A}
\end{gather}
To calculate $k_2$ we need to find the discrete quantum spectrum of the QBH.  In principle, we should solve the quantum gravity equations and find the spectrum of the BH. Remarkably, this can actually be done for specific models (see, for example, \cite{Brustein:2016msz}). Here, we rather solve the corresponding classical wave equation and then use the Bohr correspondence principle to find the spectrum in a similar way to the way that the Bohr-Sommerfeld quantization rule was used to find the spectra of atoms.  A similar procedure for  scalar waves was carried out in \cite{Brustein:2017koc}. First, we use scaling arguments to estimate $k_2$ and then support the scaling arguments by a calculation.

On dimensional grounds, the coherent state energy spectrum of macroscopic excitations of the QBH takes the classical form
$|\Delta E_{1,n_r}|\sim M_{BH}\; \omega_{n_r}^2 R^2 $, where  $\omega_{n_r}$ is the frequency of the mode $|n_r,2,0\rangle$. The matrix element of the quadrupole operator scales as
$|\langle \Psi_0|\widehat{Q}|n_r,2,0\rangle| \sim |\Delta E_{1,n_r}| R^2 \sim M_{BH}\; \omega_{n_r}^2 R^4$. It follows that each term in the sum in Eq.~(\ref{k22A}) scales as $\frac{1}{R^5}\frac{|\langle \Psi_0|\widehat{Q}|n_r,2,0\rangle|^2 }{|\Delta E_{1,n_r}|}\sim  \frac{|\Delta E_{1,n_r}|}{R} \sim \frac{|\Delta E_{1,n_r}|}{M_{BH}}\sim \omega_{n_r}^2 R^2  $. This semiclassical treatment  is supported by observing that the occupation numbers $N$, in the excited energy levels, scale as $N \hbar \omega_{n_r}\sim M_{BH}\; \omega_{n_r}^2 R^2$ so $N\sim (\omega_{n_r}R)S_{BH} \gg 1$. We may also use a scaling argument and an explicit calculation to show that the contributions to $k_2$ in Eq.~(\ref{k22A}) of the excited states above the first excited state are suppressed, so we can approximate the sum over $n_r$ by the contribution from the first excited state. This is a typical situation in most quantum systems.  Furthermore, all the terms in the sum are positive, so the approximate value of the magnitude of  $k_2$ is an underestimate.  In this case, it is justified to approximate the sum by the contribution of the first excited state.  Putting the two scaling arguments together, we get an estimate for $k_2$,
\be
k_2=~-\dfrac{3}{4 R^{5}}\dfrac{|\langle \Psi_0|\widehat{Q}|2,2,0\rangle|^2 }{|\Delta E_{1,2}|}\sim -\dfrac{|\Delta E_{1,2}|}{M_{BH}}\sim - \omega_{2}^2 R^2.
\label{k22Estimate}
\ee

We now turn to a quantitative evaluation of $k_2$, whose aim is to calculate the order unity numerical factor in Eq.~(\ref{k22Estimate}). We emphasize that the estimate in Eq.~(\ref{k22Estimate}) is valid in a model-independent way. The specific model that we discuss will serve to illustrate the procedure in a simple model for which numerical factors can be calculated analytically. Later we parametrize the Love number in terms of the single parameter $g^2$ and interpret its detectability in terms of the estimate in  Eq.~(\ref{k22Estimate}).

Because gravity in the interior of the BH is strongly coupled, one cannot use the semiclassical geometric description in terms of a curved spacetime. It needs to be replaced by describing gravity as an inertial force in a flat space, a replacement that is allowed by virtue of Einstein's equivalence principle. The specific nature of the excitations in the interior is unimportant and so is the equivalence of the two descriptions of gravity. The only relevant aspect is that excitations are macroscopic, horizon-scale excitations so that applying the Bohr principle is justified.

The idea is that the exotic matter in the interior of the QBH can be effectively
viewed as a fluid that supports pulsating modes as for a relativistic star. These fluid modes would exist in addition to the standard spacetime modes of the exterior.  The  perturbations are divided into two sectors, the fluid modes and spacetime modes. Due to their low speed of sound and the compactness of the QBH, fluid modes are decoupled from the spacetime perturbations as in the  Cowling approximation \cite{Kokkotas:1999bd,Allen:1997xj,Andersson:1996ua}.

The boundary conditions (BCs) are chosen as follows. Spherical symmetry requires fully reflecting BCs at the center of the QBH. The QBH has an outer surface that behaves just like a classical BH horizon in the classical limit. In this case,  the internal fluid modes decouple from the exterior. Then, absence of transmission, or perfect reflection at the outer surface is the correct BC.  When quantum effects are small, the outer surface is only partially opaque and so the reflection is not perfect.  We found that, quantitatively, both BCs lead to almost identical spectra. Since the analysis is much simpler in the former case, we will impose this BC at the outer surface and find the spectrum of normal modes rather than quasinormal modes.

Thus, the conclusion is that the classical equation that we need to solve  is the Laplace equation,
\be
\overrightarrow{\nabla}^2 \Psi_2(r) ~=~ 0 ,
\label{inwave}
\ee
with the generic BC
$\Psi_2{}_{|r=0}=0~$, and $\Psi_2'{}_{|r=R}=0.$
The solution of Eq.~(\ref{inwave}) is
\be
\Psi_2(r)~=~ {\cal N}_2~ j_2\left( q r\right) Y_{20}(\theta,\phi),
\label{psiin}
\ee
where $j_2$ is the spherical Bessel function, $Y_{20}$ is the (real) spherical harmonic function with $l=2$, $m=0$ and ${\cal N}_2$ is a normalization factor which will be determined later.
The BC in this case allows only discrete values on the magnitude of  the wave number $q$,
\be
j_2'\left( q R\right)~=~0,
\ee
which is very well approximated by
\be
q_{n_r}  =  \left(n_r-\tfrac{1}{2}\right) \tfrac{\pi }{R}, \hspace{.5in} n_r = 3, 4, \dots,
\label{match2}
\ee
while for $n_r=2$, the value is somewhat lower,
\be
q_2 \simeq 1.06 \tfrac{\pi }{R}.
\label{match22}
\ee

Condition (\ref{match2}) can also be viewed as a manifestation of the Bohr quantization condition in the corresponding QBH. Substituting $P=\hbar q$, we find
\be
P R ~= ~\pi \hbar (n_r-\tfrac{1}{2}).
\label{qcond}\ee

We need to calculate $|\Delta E_{1,2}|$ and
$|\langle \Psi_0|\widehat{Q}|2,2,0 \rangle|$ using the solution  $\Psi_{2,2}={\cal N}_{2,2}~ j_2\left( q_2 r\right) Y_{20}$, with the wave number given above.
First, because the classical waves are nonrelativistic,
\be
|\Delta E_{1,2}|~= ~\frac{1}{2} M_{BH} \omega_2^2 R^2 = \frac{1}{2} M_{BH} g^2 q_2^2 R^2.
\label{De1}
\ee
In the last equality, we introduced a parametrized  dispersion relation $\omega_2^2= g^2 q_2^2$, where $g^2 \ll 1$ determines the energy of the first excited level and is the only free parameter of our model. The effective index of refraction in the cavity is $1/g^2$ (see also \cite{Brustein:2017nis,Brustein:2017koc}).

To evaluate the expectation value $|\langle \Psi_0|\widehat{Q}|2,2,0 \rangle|$, Eq.~(\ref{k22Estimate}), we will need a  more elaborate calculation.
First, we need the general expression for the excitation energies for $n_r\ge 3$,
\be
|\Delta E_{1,n_r}|~=~ \frac{1}{2} g^2 M_{BH} \pi^2(n_r-\tfrac{1}{2})^2,
\label{De1}
\ee
where we have absorbed any additional $n_r$-independent factors into $g^2$ and assumed that the dispersion relation is the same for all modes. The excitation energy has to be parametrically small compared to the BH mass, $|\Delta E_{1,n_r}|\ll M_{BH}$. This condition restricts the validity of the estimate in Eq.~(\ref{De1}) and the range of $n_r$ in the sum in Eq.~(\ref{k22A}) (see also the discussion in the subsequent section).

 To proceed, the classical quantity that corresponds to the matrix element $|\langle \Psi_0|\widehat{Q}|2,2,0 \rangle|$ is given by
\be
|\langle \Psi_0|\widehat{Q}|2,2,0 \rangle|\!\!~\leftrightarrow~\!\! \int r^2 dr\; d\Omega_2   \Delta\rho_{2,2}(r) r^2\;  Y_{20}\; \Psi_{2,2}~.\;
\label{matrixCorr}
\ee
This quantity is evaluated by calculating the effective energy density in the first excited state, $\Delta\rho_{2,2}(r)$, using the following comparison.
On one hand,
\be
|\Delta E_{1,n_r}|~=~\int r^2 dr\; \Delta\rho_{2,n_r}(r).
\ee
On the other hand, to lowest order in $g^2$, the energy $|\Delta E_{1,n_r}|$ is proportional to $\omega_{n_r}^2$,
\bea
|\Delta E_{1,n_r}|\!~&=&~\! \int r^2 dr\; d\Omega_2 \; |\Psi_{2,n_r}|^2 \omega_{n_r}^2
\cr \!&=&\!
~|{\cal N}_{2,n_r}|^2 \int_0^R r^2 dr\;  j_{2}^2 \left(\tfrac{\omega_{n_r}}{g} r\right)   \omega_{n_r}^2,
\label{dE1}
\eea
where we have used Eq.~(\ref{psiin}) and Eq.~(\ref{match2}) and performed the angular integral.
Comparing the two expressions for $|\Delta E_{1,n_r}|$, we find that
\be
\Delta\rho(r)_{2,n_r}(r) ~=~ \frac{|\Delta E_{1,n_r}|}{I_{2,n_r}} j_{2}^2 \left(\tfrac{\omega_{n_r}}{g} r\right)
\label{dE2}
\ee
and
\be
|{\cal N}_{2,n_r}|^2~=~  \frac{|\Delta E_{1,n_r}|}{ \omega_{n_r}^2 I_{2,n_r}},
\label{normalization2}
\ee
where $I_{2,n_r}=\frac{R^3}{\pi^3 \left(n_r-\frac{1}{2}\right)^3}\int\limits_0^{\pi(n_r-\frac{1}{2})}\!\!\!  y^2 dy\; j_2^2(y)$.
Substituting Eq.~(\ref{dE2}) into expression (\ref{matrixCorr}) results in the following expression:
\bea
|\langle \Psi_0|\widehat{Q}|n_r,2,0 \rangle|&\leftrightarrow &
\int r^2 dr\; d\Omega_2\;  \frac{|\Delta E_{1,n_r}|}{I_{2,n_r}}\; j_{2}^3 \left(\tfrac{\omega_{n_r}}{g} r\right)   (Y_{20})^2
\cr &=&
|\Delta E_{1,n_r}|\; {\cal N}_{2,n_r}\frac{I_{4,n_r}}{I_{2,n_r}},
\label{qmatrix}
\eea
where $I_{4,n_r}=\frac{R^5}{\pi^5 \left(n_r-\frac{1}{2}\right)^5}\int\limits_0^{\pi(n_r-\frac{1}{2})} dy y^4 j_2^3(y)$.

Putting  all the pieces together we find that the corresponding expression to the ratio appearing in Eq.~(\ref{k22A}) is the following:
\bea
\frac{\langle|\Psi_0|\widehat{Q}|n_r,2,0\rangle|^2 }{|\Delta E_{1,n_r}|} ~\leftrightarrow~
\frac{|\Delta E_{1,n_r}|^2}{ \omega_{n_r}^2} \frac{I_{4,n_r}^2}{I_{2,n_r}^3}.
\label{almostk2}
\eea

The sum  of terms with $n_r\ge 3$ in Eq.~(\ref{k22A}) is therefore given by
\bea
&&\sum_{n_r=3}  \frac{|\Delta E_{1,n_r}|^2}{\omega_{n_r}^2} \frac{I_{4,n_r}^2}{I_{2,n_r}^3}
~\\&=&~\frac{1}{4}g^2 M_{BH}^2 R^3 \sum_{n_r=3} \pi(n_r-\tfrac{1}{2})  \left(\widetilde{I}_{4,n_r}\right)^2\;\left(\widetilde{I}_{2,n_r}\right)^{-3}, \nonumber	\label{k22B}
\eea
where we also use the energy spectrum Eq.~(\ref{De1})
and the integral $\widetilde{I}_{2,n_r}=\int\limits_0^{\pi(n_r-\frac{1}{2})}\!\!\!  y^2 dy\; j_2^2(y)$  scales linearly with $\pi(n_r-\tfrac{1}{2})$, and the integral $\widetilde{I}_{4,n_r}=\int\limits_0^{\pi(n_r-\frac{1}{2})}\!\!\!  y^4 dy\; j_2^3(y)$ is approximately a constant. The different scalings arise because of the different scaling of integrals of even and odd powers of the spherical Bessel function.  The final result is that the terms in the sum scale as $1/(\pi(n_r-\tfrac{1}{2})^2$, with odd $n_r$ terms being much smaller than even $n_r$ terms. The $n_r=2$ term is the largest in the sum and next largest term is the $n_r=4$ term, whose magnitude is about 1/5 of the $n_r=2$ term.

Once both $|\Delta E_{1,2}|$ and $|\langle \Psi_0|\widehat{Q}|2,2,0 \rangle|$ are known, they can be substituted into Eq.~(\ref{k22Estimate}). The result is given by
\begin{gather}
	k_2~=~
	-\dfrac{3}{16} \dfrac{1}{q_2 R} \frac{M_{BH}^2}{R^2}\frac{ \widetilde{J}_4^2}{\widetilde{J}_2^3}~ \omega_2^2 R^2 \cr
	~=~ -\dfrac{3}{16} {q_2 R} \frac{M_{BH}^2}{R^2}\frac{ \widetilde{J}_4^2}{\widetilde{J}_2^3} \; g^2,
	\label{k22Single}
\end{gather}
where the integrals $\widetilde{J}_{2}=\int\limits_0^{q_2 R} y^2 dy\;j_2^2(y)$ and  $\widetilde{J}_{4}=\int\limits_0^{q_2 R} y^4 dy\;j_2^3(y)$ can be evaluated analytically. Substituting the numerical values of the integrals and setting $M_{BH}/R=1/2$, we  arrive at our final result,
\bea
k_2&=&-0.09~ \omega_2^2 R^2=-0.18~ \frac{|\Delta E_{1,2}|}{M_{BH}} \cr
&=& -0.99~ g^2.
\label{k22C}
\eea
As anticipated in Eq.~(\ref{k22Estimate}), $k_2$ scales as $ \omega_2^2 R^2$.

We can compare the value of $k_2$ in Eq.~(\ref{k22C}) to the values of $k_2$ for other compact objects.  For Neutron stars $k_2$ is positive and its magnitude is much larger than the value of $k_2$ in Eq.~(\ref{k22C}).
For the exotic UCOs, universal logarithmic dependence was found in \cite{Cardoso:2017cfl}. These analyses assumed that some modifications lead to a shift at the UCO outer surface $R=2M(1+\epsilon)$ and concluded  $k_2\sim 1/|\ln\epsilon|$ and that it is negative.  The real part of the frequency of spacetime modes for these UCOs for the $n=2$ mode is $\omega_{{2,UCO}}\sim {1/ |\ln\epsilon|}$ \cite{Maggio:2018ivz}.  For the BH area quantization model \cite{Bekenstein:1995ju} (see also \cite{Maggiore:2007nq,Foit:2016uxn,Cardoso:2019apo,Agullo:2020hxe}), $\omega_n={  \alpha n}/{16\pi R}$, with  $\alpha$ being a dimensionless coefficient of order unity, so we can apply our semiclassical treatment and from Eq.~(\ref{k22Estimate}) calculate  the Love number $k_2\simeq  ~\frac{3}{16} \left(\frac{\alpha}{8 \pi}\right)^2$.

\end{subequations}

\section{Detectability}\label{dec}
\begin{subequations}
Here, we discuss the possibility of measuring the quantum Love number in future LISA observations of supermassive BH binaries, which LISA can observe from the early stages of the inspiral up to the coalescence. We show that for such binary systems,  the sensitivity is sufficient for possibly detecting the quantum tidal deformation effects for a range of values of $g^2$. We include here also the case of moderately spinning BHs whose dimensionless spin parameter is $\chi\lesssim 0.7$. We later show that the main effect of the spin is to modify the radius of the BH for the same mass, $R= M(1+\sqrt{1-\chi_i^2})$ and that the direct effect of the spin on the spectrum of the BH can be neglected.

Following  \cite{Maselli:2017cmm,Maselli:2018fay} (see also \cite{Cutler:1994ys,Yagi:2013baa,Sennett:2017etc}), we determine for which values of $g^2$, the statistical error due to the detector noise is small enough for observing the tidal deformation effects.   We also need to include tidal heating effects \cite{Poisson:1994yf,Alvi:2001mx,Poisson:2004cw,Taylor:2008xy} which are present because QBHs posses a horizon. However, we found that these induce small changes to the error estimation.

To estimate the statistical error in measuring the Love number, we use
a parameter estimation method based on the Fisher  matrix
$\Gamma_{ij}=(\frac{\partial h}{ \partial \theta^i}|(\frac{\partial h}{ \partial \theta^j})$, where the inner product $(\cdot|\cdot)$ is defined by
$
(h_1|h_2)~=~4~\text{Re}\int _{f_{min}}^{f_{max}} \frac{\tilde{h}_1(f)\tilde{h}_2^*(f)}{S_n(f)}df~.\label{hhi}
$
The LISA noise spectral density is denoted by $S_n(f)$ \cite{LISA,Cornish:2018dyw}.  The minimal frequency of LISA's observation band is denoted by $f_{min}$, $f_{min}\approx 10^{-5}$ Hz which corresponds to an observation time of about one year \cite{Berti:2004bd}. The maximal frequency $f_{max}$ is taken to be the frequency at the innermost stable circular orbit (ISCO) \cite{Favata:2010ic}.
The model signal and the true signal are parametrized by the function  $\theta^i=(\text{ln}~\mathcal{A},\text{ln} ~\mathcal{M},\text{ln} ~\eta, \Psi_c,t_c,\chi_1,\chi_2, \Lambda)$, whose arguments are the amplitude $\mathcal{A}$, the chirp mass $\mathcal{M}=\eta^{3/5}M$, the symmetric mass ratio $\eta=M_1M_2/M^2$, the phase $\Psi_c$, the time at coalescence $t_c$, the dimensionless spin parameters  $\chi_1,\chi_2$ and the
dimensionless average tidal deformability parameter
$
\Lambda=\frac{16}{13}\left[\left(1+12\frac{M_2}{M_1}\right)\frac{M_1^5}{M^5}\widetilde{\Lambda}_1+ \left(1+12\frac{M_1}{M_2}\right)\frac{M_2^5}{M^5}\widetilde{\Lambda}_2\right],
$
where $M=M_1+M_2$, $\widetilde{\Lambda}_i=\lambda_i/M^5$ and $\lambda_i$ is defined in Sec.~\ref{int}.
For this set of parameters, the root-mean-square error in measuring $\Lambda$ is expressed through the inverse of the Fisher matrix
$
\sigma_\Lambda=\sqrt{\langle \left(\Delta \Lambda\right)^2\rangle}= \sqrt{\left(\Gamma^{-1}\right)_{\Lambda \Lambda}}~.
$

For a  binary inspiral, the Fourier transform of the signal is modeled by  $\tilde{h}(f,\theta^{i})=\mathcal{A} e^{i \Psi}$, where $\Psi=\Psi_{PP}+\Psi_{TD}+\Psi_{TH}$ are the phases of the point-particle, tidal deformability and tidal heating effects,  respectively.

The approximation method adopted here is the analytical ``TaylorF2 approximant'' \cite{Damour:2000zb,Arun:2004hn,Buonanno:2009zt}. We include correction terms to the GW phase in the form of spin-orbit, spin-spin and cubic-spin corrections up to 3.5 PN order relative to the leading-order GW term \cite{Isoyama:2017tbp,Khan:2015jqa},  tidal deformability terms to 5 PN and 6 PN order \cite{Sennett:2017etc,Bini:2012gu,Hotokezaka:2016bzh}, and tidal heating correction term for spinning BHs to the leading 2.5 PN order relative to the leading-order GW term \cite{Isoyama:2017tbp,Cardoso:2019rvt}. The amplitude is taken to leading PN order and includes the sky-averaged prefactor \cite{Berti:2004bd}.
\begin{figure}[h!]
	\centering
\hspace{-0.6cm}	\includegraphics[width=1\linewidth]{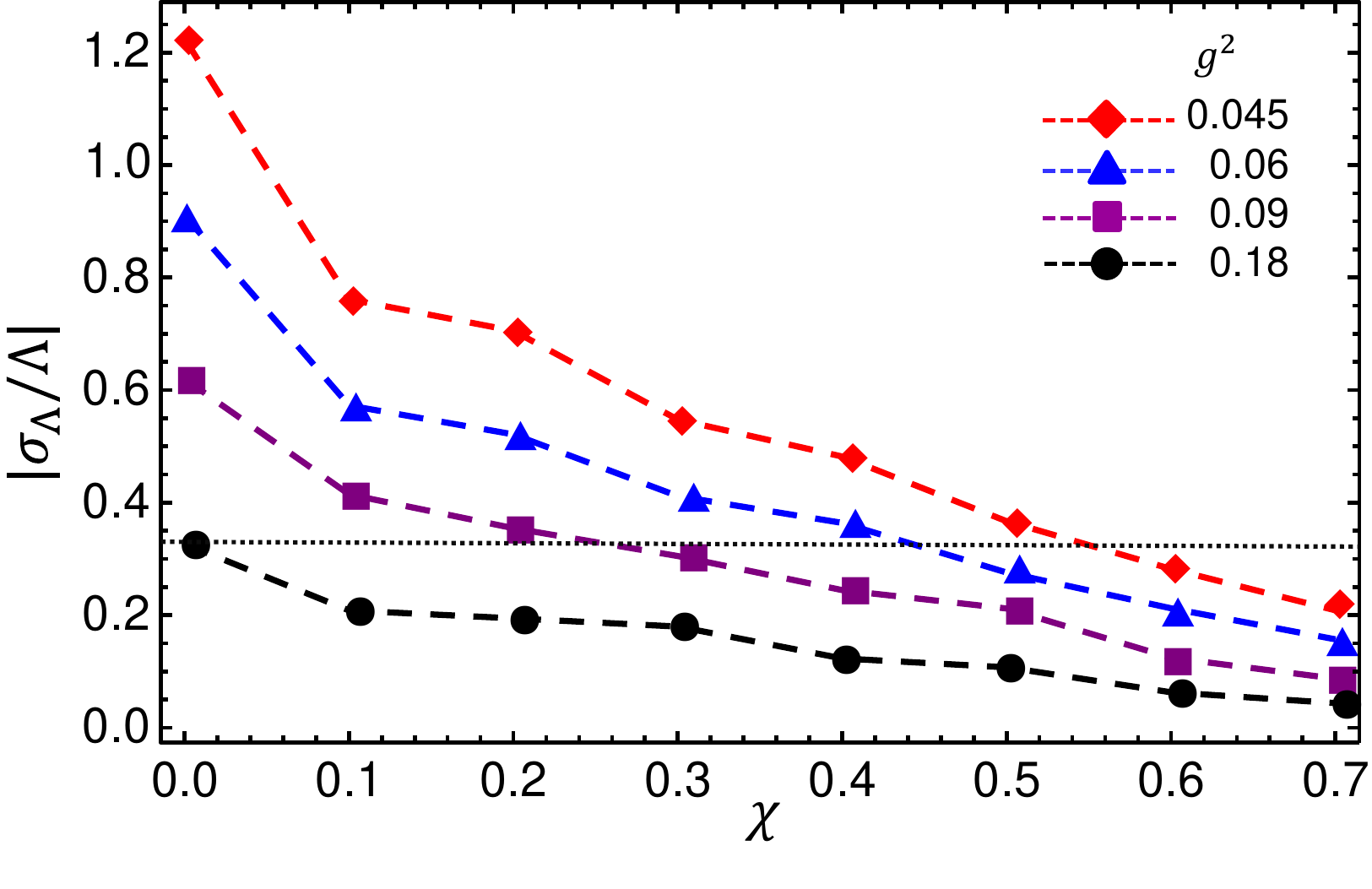}
\vspace{-0.3cm}	\caption{{\footnotesize The relative statistical error in future LISA observations of $|\sigma_\Lambda/\Lambda|$  for several values of $g^2$ is shown  as a function of the spin parameter $\chi$. We assume equal spin and equal mass $M=10^6 M_{\odot}$ companions and that the luminosity distance to the binary system is $D_l=2$ Gpc. Points below the horizontal dashed line correspond to detections at the 3$\sigma$ level.  The value $g^2=0.18$ is a limiting value for which $|\Delta E_{1,2}|= M_{BH}$.}}
	\label{s2}
\end{figure}

The justification for using the TaylorF2 approximate to estimate the detectability of the Love number of the QBHs is the following.  As previously discussed, the only difference between a BH and a QBH is in the response of the QBH to external perturbations.  Except for this difference, BHs and QBHs are indistinguishable to a distant external observer as both can be viewed as point masses, being well described by the spherically symmetric vacuum solution. Thus, the TaylorF2 approximation for the QBH and the BH is identical up to the subleading 5th PN order in which tidal deformation effects enter. Consequently, the use of the TaylorF2 approximate for BH-like objects is a standard accepted practice in similar contexts\cite {Yagi:2013baa,Cardoso:2017cfl,Maselli:2017cmm}.

The results presented in Fig.\ref{s2} indicate that it is possible to place significant constraints on, or possibly measure the quantum Love number,  $|k_2|=3.96\times g^2(1+\sqrt{1-\chi_i^2})^{-2}$, for supermassive, moderately spinning binaries ($M\sim 10^6 M_{\odot},~ \chi_i\lesssim 0.7$) at luminosity distance $D_l=2$ Gpc. For example, taking $g^2=0.06$ and for spin $\chi=0.6$ (so $k_2\approx 0.073$), the relative error $\left|\sigma_\Lambda/\Lambda\right| \approx 0.2$ leads to detections at $5\sigma$ confidence. Our results suggest that it would be possible to measure $\alpha\simeq 12$ of the area quantization model to better than 1$\sigma$ confidence.

We wish to emphasize that the effect of degeneracy among the parameters could have been important for determination of the statistical error on $\Lambda$. However,  as pointed out in \cite{marko,Hughes:2001ya}, even when the degeneracy is maximized, its effect would have increased the relative error on $\Lambda$ by not more than its square root. We conclude that including the effects of degeneracy is not required at the level of accuracy that we have adopted, as it would not have changed  our main conclusion significantly.
\end{subequations}
\begin{subequations}

	\renewcommand{\theequation}{\theparentequation.\arabic{equation}}

\section{Summary}

In this paper we calculated the Love number of QBHs using standard time-independent
quantum perturbation theory. We showed that, unlike classical BHs whose Love numbers vanish, the Love numbers of QBHs are generically nonvanishing and negative and their magnitude depends most strongly on the first excited level of the quantum spectrum. We focused on evaluating the largest Love number $k_2$, the electric quadrupolar Love number.
Replacing quantum expectation values  by the corresponding classical quantities, as dictated by the Bohr correspondence principle, we found that $k_2$ of nonrotating QBHs takes the universal form
\begin{gather}
k_2~=~-\mathcal{N}\omega_2^2R^2	
\label{k2n}~,
\end{gather}
where $\mathcal{N}$ is a positive numerical factor of order unity that is determined by the generic boundary conditions of QBHs, Eqs.~(\ref{inwave}) and (\ref{psiin}), and the object's excitation spectrum. As shown in Sec.\ref{calc}, the result in Eq.(\ref{k2n}) is universal and holds for any  macroscopic quantum object.

We then proceeded to show that the accumulated dephasing due to the dissipation of tidal deformation in supermassive moderately spinning binaries during $\sim1$year of observation is large enough to induce a significant deviation on the orbital phase. Thus, indicating the detectability of the Love number of QBHs with future precision GW measurements.

\end{subequations}
\section*{Acknowledgments}
We thank Zvi Bern, Vitor Cardoso and  Julio Parra-Martinez for discussions and Kent Yagi for comments on the manuscript.
The research of R. B. and Y. S. was supported by the Israel Science Foundation Grant No. 1294/16. The research of Y. S. was supported by the Negev scholarship. R. B. and Y. S. thank the Theoretical Physics Department, CERN for their  hospitality.

\appendix
\section{Effects of spin}

In this appendix, we discuss two effects that depend on  the spin of the BH. First,  the recent discovery in \cite{Poisson:2020mdi} that magnetic Love numbers of a slowly rotating object should be promoted to tensors and second, we show that the direct effect of the spin on the spectrum of the BH can be neglected, thus justifying the statement in the text that the main effect of the spin is to modify the radius of the BH for the same mass, $R= M(1+\sqrt{1-\chi_i^2})$.

Recently, in \cite{Poisson:2020mdi}, it was demonstrated that the magnetic Love numbers of a slowly rotating object should be promoted to tensors. For example, following \cite{Poisson:2020mdi}, the magnetic Love tensor for  $l=2$  is given by
$(k_{2})^{ij}_{~~kl}=(k^{B}_{2})^{ij}_{~~kl}+(k^{M}_{2})^{ij}_{~~kl}$. The scalar component $(k^{B}_{2})^{ij}_{~~kl}$ is related to the magnetic Love number given in Eq.~(\ref{magk}),
$(k_2^B)^{ij}_{~~kl}=k_2^B \delta_k^i \delta_l^j$. The additional spin induced term is given by
\begin{eqnarray}
	(k^{M}_{2})^{ij}_{~~kl}\!\!=-\sum_{n>1,|m|\leq 2}\dfrac{3}{4R^{5}}~ \dfrac{|\langle \Psi_0|\widehat{S}^M_{~~ij}|n_r,~2,~m\rangle|^2 }{\Delta E_{1,n_r}}\times  \mathcal{N}(\mathcal{Y}_2^m)^{ij}(\mathcal{Y}_2^m)_{kl},
\end{eqnarray}
where $\hat{S}_{~~ij}^M$ is the $l=2$  current moment, $(\mathcal{Y}_2^m)_{kl}$ are azimuthal symmetric-free tensors and $\mathcal{N}$ is a numerical factor that is determined by the orthogonality of the generalized spherical functions (see definitions in \cite{Poisson:2020mdi}). Similarly the magnetic Love tensors for a general $l$ can be obtained.

When the BH is spinning, its spin is coupled to the orbital tidal field. The interaction energy takes the form \cite{Pani:2015nua,Abdelsalhin:2018reg}
\begin{gather}
	V_{int}~=~ -Q_{ij}\mathcal{E}^{ij}\cr ~=-
	\lambda_2\left(\mathcal{E}_{ij}+2 \alpha \mathcal{B}_{ijk}J^k/M\right)\mathcal{E}^{ij},
\end{gather}
where $J^k=M^2\chi n^k$ is the spin vector ($ \chi $ is the dimensionless spin parameter), $\alpha$ is a dimensionless coefficient of order unity or less \cite{Pani:2015nua} and $\mathcal{B}_{ijk}$ is the $l=3$ octupolar tidal field. In this form, since $|\mathcal{B}_{ijk}|\sim M |\mathcal{E}_{ij}|v^3$ it is clear that the spin corrections are 1.5PN order higher than the leading quadrupolar term, and therefore can be neglected.

One can also view this as spin corrections to the quadrupole moment, $\delta Q_{ij}=-\lambda_2 2 \alpha \mathcal{B}_{ijk}J^k$, or as spin corrections to the tidal Love number,
\begin{gather}
	\lambda_2 ~\sim~\lambda_2^{\chi=0}\left(1+2\alpha \left|\dfrac{\mathcal{B}_{ijk}}{\mathcal{E}_{ij}}\right|\dfrac{\chi}{M}\right) \sim \lambda_2^{\chi=0}\left(1+ v^3 \chi\right).
\end{gather}

\end{document}